\documentclass[twocolumn,prb,showpacs]{revtex4-1}
\usepackage{graphicx}    
\usepackage{dcolumn}     
\usepackage{bm}          
\usepackage{amsmath}
\newcommand{\ave}[1]{\left \langle #1 \right \rangle}
\newcommand{\abs}[1]{\left| #1 \right|}
\begin{document}
\title{Asymmetry of Superconductivity in Hole- and Electron-doped Cuprates:\\ Explanation within Two-Particle Self-Consistent Analysis for the Three-Band Model}
\author{Daisuke Ogura and Kazuhiko Kuroki}
\affiliation{Department of Physics, 
Osaka University, 1-1 Machikaneyama, 
Toyonaka, Osaka 560-0043, Japan}
\begin{abstract}
In the hole-doped cuprate superconductors, the superconducting transition temperature $T_c$ exhibits a dome-like feature against the doping rate. By contrast, recent experiments reveal that $T_c$ in the electron-doped systems monotonically increases as the doping is reduced, at least up to a very small doping rate.
Here we show that this asymmetry is reproduced by performing a two-particle self-consistent analysis for the three-band model of the CuO$_2$ plane. This is  explained as a combined effect of the intrinsic electron-hole asymmetry in systems comprising Cu3$d$ and O2$p$ orbitals and the band-filling-dependent vertex correction. 
\end{abstract}
\maketitle
Despite the long history, there still remain various unsolved problems in the study of the high-$T_c$ cuprate superconductors. 
The striking difference in the doping dependence of the superconducting transition temperature $T_c$ 
between the hole- and the electron-doped materials is among those unresolved issues. It is well known that in the hole-doped case, $T_c$ exhibits a dome-like feature against the doping rate, namely, $T_c$ first increases upon doping (underdoped), then yields a maximum value (optimal), and finally decreases with further doping (overdoped). On the other hand, it was known for the electron-doped cases that $T_c$ abruptly appears as soon as the antiferromagnetism is lost with doping, and monotonically decreases as the doping rate increases. Recent experiments show that the antiferromagnetism can be suppressed down to very small doping rate, or even in the mother compound, when the apical oxygens are ideally removed in the T$^\prime$-type crystal structure of the electron doped cuprates. Then, it has been revealed that $T_c$ monotonically increases  with decreasing the electron doping at least up to a very small doping rate (less than 5 percent), and is suggested to be superconducting even in the non-doped mother compound\cite{Tsukada2005,Brinkmann1995,Krockenberger2013,Adachi2013}.

There have been some theoretical studies of the doping dependence of $T_c$. 
The fluctuation exchange (FLEX) approximation\cite{Bickers1989} for the single band Hubbard model gives a monotonic doping dependence of $T_c$\cite{Bickers1991}, and therefore has difficulties in understanding the doping dependence of $T_c$  in the hole-doped cuprates. Some studies considered superconducting fluctuation in FLEX to circumvent this problem\cite{Yanase2001,Kobayashi2001}. There have also been some studies  that adopt methods capable of dealing with the strong correlation effects \cite{Gull2013,Kancharla_Kotliar_Tremblay2008,Senechal_Tremblay,Yokoyama2013,Jarrell2001,Kyung_Tremblay2003,Chen2013}. 
In some of those studies, $T_c$ exhibits a dome-like doping dependence, but in those cases there would be difficulties in understanding the recent experimental results for the electron-doped case. The electron-hole asymmetry of $T_c$ was studied in a two band model that explicitly considers the O$2p$ orbital, but there, the antiferromagetic phase was obtained in a wide electron doping range\cite{Kobayashi2002}, in contradiction to the experiments mentioned above\cite{Tsukada2005,Brinkmann1995,Krockenberger2013,Adachi2013}.

It has been suggested that the difference in the character of the mother compound (Mott insulator or not) between the hole-doped and the  electron-doped systems can be attributed to the difference in the electronic structure originating from the crystal structure\cite{Yokoyama2013,Weber2010,Adachi2013,Das2009}.
Namely, while the crystal structure of the single-layer hole-doped cuprates is composed of Cu-O octahedra (T-type),
that of the electron-doped cuprates is composed of Cu-O squares (T$^\prime$-type) and (ideally) has no apical oxygens.
Due to this difference, the copper 3$d$ -oxygen 2$p$
level offset in the T$^\prime$-type structure tends to be smaller than that in the T-structure. Since the $d$-$p$ level offset is small in the electron-doped system,the on-site effective $U$, when mapped to the single-band Hubbard model, is also small. One might expect that this difference in the crystal structure, and hence the difference in the effective on-site $U$,  can provide an explanation for the electron-hole doping asymmetry of $T_c$. However, the inner layers of multi-layered hole-doped cuprates also do not have apical oxygens and therefore have the same lattice structure as that of the electron doped cuprates. Still, it is known that $T_c$ exhibits a dome-like doping dependence even within the inner layers\cite{Mukuda2012}. Therefore, it seems difficult to attribute the electron-hole asymmetry of the doping dependence of $T_c$ to the absence/presence of the apical oxygens. The aim of the present study is to understand the origin of this electron-hole asymmetry. Here, we 
stress that in the present study we focus only on the (non-)monotonicity of the doping dependence of $T_c$, and leave the issue of the metallicity or Mottness of the mother compound  to future studies.

We start by demonstrating that this electron-hole 
doping asymmetry of $T_c$ is difficult to understand within the single band 
Hubbard model even when realistic band structures are considered.
We perform first principles band calculation of HgBa$_2$CuO$_4$ (a hole-doped system) and Nd$_2$CuO$_4$ (an electron-doped system),
and obtain tight-binding models constructing maximally-localized Wannier basis\cite{Blaha2001wien2k,Kunevs2010wien2wannier,Krasse1993VASP,Krasse1999VASP,Mostofi2008wannier90}.
Instead of the typical T-type hole doped system La$_2$CuO$_4$, we adopt HgBa$_2$CuO$_4$ because (i) it is known that the hybridization of the $d_{z^2}$ orbital cannot be neglected in La$_2$CuO$_4$\cite{Sakakibara2010}, and (ii) the band structures of HgBa$_2$CuO$_4$ and Nd$_2$CuO$_4$ are very similar, so that we can concentrate purely on the electron-hole asymmetry.
To take into account the electron correlation effect beyond those taken into account in the LDA/GGA level, the on-site interaction has to be treated 
by some many-body technique as has been done in previous 
studies\cite{Sakakibara2010,KotliarRMP2006,Suzuki2014}.
In the present study, we adopt the two-particle self-consistent method (TPSC) proposed by Vilk and Tremblay \cite{Vilk1997}.
In this method, the interaction vertices in the charge and spin channel are approximated as different constants, and these constants 
are determined so that the correlation functions satisfy their sum rules that originates from the Pauli's principle.
It has been shown in ref.
\cite{Kyung_Tremblay2003} that TPSC 
gives a dome-like doping dependence of $T_c$ for the single band Hubbard model 
with  nearest neighbor hopping only. 

\begin{table}[tb]
\centering
\caption{\label{table1} Nearest ($t_1$), second ($t_2$) and third ($t_3$) 
neighbor hopping integrals for the single-band models.}
\begin{ruledtabular}
\begin{tabular}{c ccc}
 & $t_1$(eV)  & $t_2$(eV)  & $t_{3}$(eV)   \\ \hline
Nd$_2$CuO$_4$ & -0.457& 0.0866 & -0.0865\\
HgBa$_2$CuO$_4$ & -0.464 & 0.0907 & -0.0842 \\
\end{tabular}
\end{ruledtabular}
\end{table}
\begin{table}[t]
\centering
\caption{\label{table2}Hopping integrals and copper 3$d$ -oxygen 2$p$ level offset in the three-band models.}
\begin{ruledtabular}
\begin{tabular}{c cccc}
 & $t_{dp}$(eV) & $t_{p_x p_y}$(eV) & $t_{p_x p_x}$(eV) &$\Delta_{dp}$(eV)   \\ \hline
Nd$_2$CuO$_4$ & 1.18 & -0.621 & 0.137 & 1.83 \\
HgBa$_2$CuO$_4$ & 1.26 & -0.632 & 0.133 & 2.06 \\
La$_2$CuO$_4$ & 1.38 & -0.616 & 0.0899 & 2.73
\end{tabular}
\end{ruledtabular}
\end{table}

\begin{figure}[t]
\centering
\includegraphics[width=8.0cm,clip,bb=0 0 550 550]{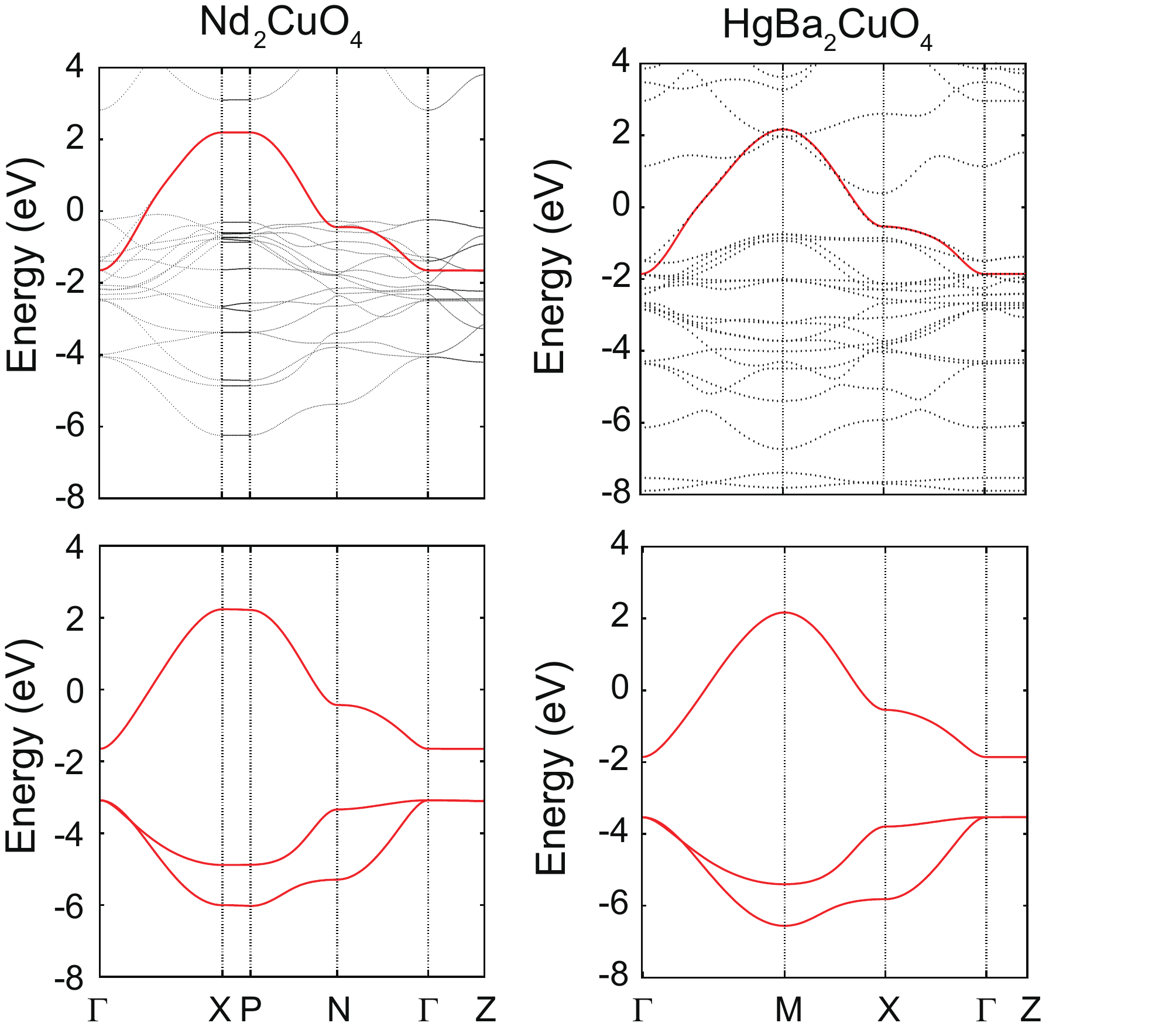}
\caption{(color online). Band structure of the single-(upper panels) and three-band models(lower) for Nd$_2$CuO$_4$(left) and HgBa$_2$CuO$_4$(right).}
\label{fig1}
\end{figure}

The obtained hopping integrals for the single-band models are given in Table \ref{table1}, and the corresponding band structures are shown in Fig.\ref{fig1}(upper panels).
The eigenvalue $\lambda$ of the linearized Eliashberg equation for $d$-wave pairing, which is a measure of $T_c$ (see below), 
is shown against the band filling in Fig.\ref{fig2}.
For both the models of HgBa$_2$CuO$_4$ and Nd$_2$CuO$_4$, we set the on-site repulsion as $U/t = 8$ and the temperature $T/t = 0.08$,
and we take 128$\times$128 meshes and 4096 Matsubara frequencies.
As shown in Fig.\ref{fig2}, $\lambda$ varies monotonically in both the hole- and the electron-doped cases, namely, the dome-like $T_c$ variance against doping  obtained for the nearest-neighbor-hopping-only case (inset of Fig.\ref{fig2} shows the doping dependence of $\lambda$ for the $t_1$-only model) is lost when a realistic band structure is adopted. 

\begin{figure}[t]
\centering
\includegraphics[width=8.0cm,clip,bb=0 0 350 300]{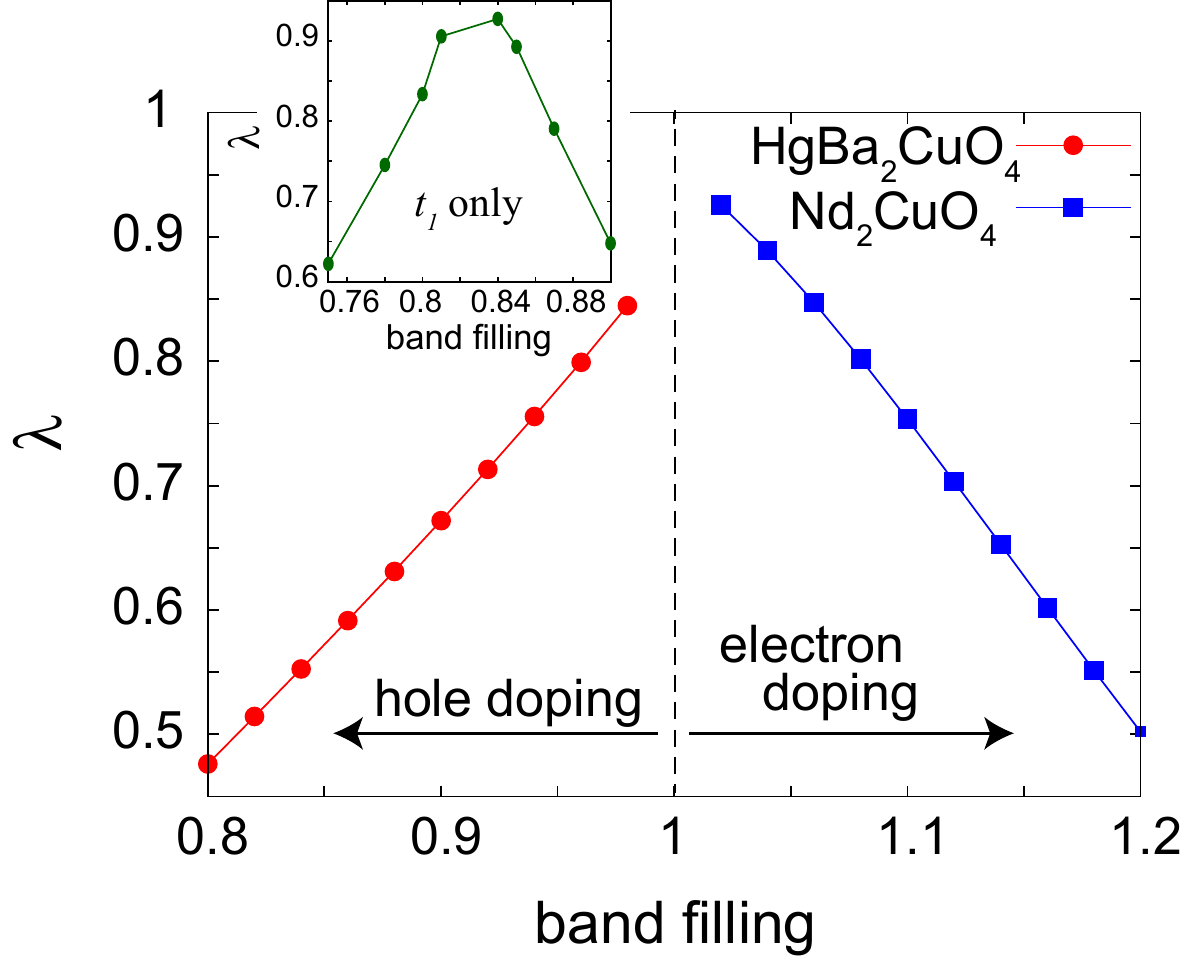}
\caption{(color online). Band filling dependence of the eigenvalue $\lambda$ of the linearized Eliashberg equation for the single band models.
For comparison,	the inset shows the doping dependence of $\lambda$ for the single band model with nearest neighbor hopping only.}
\label{fig2}
\end{figure}

Considering the previous studies mentioned in the introduction, it may be questionable whether we can reproduce the experimentally observed electron-hole asymmetry within the single band model even if we take into account the electron correlation effects beyond TPSC.  Namely, the doping dependence of $T_c$ would be either dome-like shaped or monotonic on both electron and hole-doped cases 
when the same values of $U$ are taken. Hence, we now proceed to the three-band model that explicitly considers the in-plane oxygen 2$p_{x,y}$ orbitals in addition to the copper 3$d_{x^2-y^2}$ orbital \cite{Emery1987}.
We first constructed the five-band model composed of the copper 3$d_{x^2-y^2}$ orbital and four in-plane oxygen 2$p_{x,y}$ orbitals by using the maximally-localized Wannier basis. Subsequently, we obtained the three-band model by removing two $p_\pi$ orbitals which are oriented 
in the direction perpendicular to the Cu-O bond.
The obtained model parameters and the band structure are shown in Table \ref{table2} and Fig.\ref{fig1}, respectively.
For comparison, the model parameters for La$_2$CuO$_4$ are also shown.
The parameter values of Hg and Nd systems can be considered as quite similar, and especially the similarity of $\Delta_{dp}$ can be 
noticed if we compare the values to that of the La system, which has smaller apical oxygen height compared to the Hg system. The similarity of $\Delta_{dp}$ between the electron-doped and the hole-doped materials is expected to become even more prominent if we consider  
multilayer hole-doped cuprates, where one or both of the apical oxygens are missing depending on the layer. This means similar values of 
the on-site $U$ when mapped to single band Hubbard models, as mentioned in the introductory part.

To analyze the three-band model, the TPSC approach should be generalized for multi-band systems.
We follow the generalization of TPSC presented in 
Refs.\cite{Miyahara2013,Arya2015}.
Let us briefly review TPSC for the multi-band Hubbard model.
Hereafter we make use of the matrix form in the same way as Refs.\cite{Takimoto2004,Mochizuki2005}.

The Hamiltonian of the three-band model is given as 
\begin{eqnarray}
H =&& \sum_{\bm{r},\bm{r}^{\prime},\sigma} \sum_{\mu,\nu} t_{\bm{r} \bm{r}^{\prime}}^{\mu \nu} c_{\mu\sigma}^{\dagger}(\bm{r}) c_{\nu\sigma}(\bm{r}^{\prime})
+\Delta_{dp}\sum_{\bm{r},\sigma} n_{d\sigma}(\bm{r})\nonumber \\ 
&&\quad+\sum_{\bm{r}} \sum_{\mu} U_{\mu} n_{\mu\uparrow}(\bm{r}) n_{\mu\downarrow}(\bm{r}) ,\samepage
\end{eqnarray}
where $c^{\dag}_{\mu\sigma}(\bm{r})$ is a creation operator of an electron with spin $\sigma$ and orbital $\mu=d, p_{x}, p_{y}$
at site $\bm{r}$, $n_{\mu\sigma}(\bm{r})=c_{\mu\sigma}^{\dagger}(\bm{r})c_{\mu\sigma}(\bm{r})$ is a number operator, 
$\Delta_{dp}$ is the $d$-$p$ level difference, $U_\mu$ is the on-site Coulomb interaction. The band filling $n$ is defined as the average number of electrons per unit cell, so that $n=5$ corresponds to the non-doped case. 
In the analysis for this model, we set the on-site interaction $U_d=10$eV and $U_p=5$eV, temperature $T=0.01$eV.
We employ 64$\times$64 $k$-point meshes and 4096 Matsubara frequencies.

In this three-band model, similar to the single-orbital case, the spin and charge susceptibilities are evaluated as
\begin{subequations}
\label{susceptibilities}
\begin{eqnarray}
\bm{\chi}^{\rm{sp}}(q) &=& \left[\bm{1} - \bm{\chi}^{0}(q) \bm{U}^{\rm{sp}}\right]^{-1}\bm{\chi}^{0}(q) , \label{spin suscep} \\
\bm{\chi}^{\rm{ch}}(q) &=& \left[\bm{1} + \bm{\chi}^{0}(q) \bm{U}^{\rm{ch}}\right]^{-1}\bm{\chi}^{0}(q) , \label{charge suscep}
\end{eqnarray}
\end{subequations}
where $\bm{\chi}^{0}(k)$ is the irreducible susceptibility and $\bm{U}^{\rm{sp(ch)}}$ is the effective interaction matrix for the spin (charge) channel.
The irreducible susceptibility is given by
\begin{equation}
\chi^{0}_{\lambda\mu\nu\xi}(q)=-\frac{T}{N}\sum_{k} G^{0}_{\nu \lambda}(k) G^{0}_{\mu\xi}(k+q) ,
\end{equation}
using the bare Green's function $G^{0}_{\mu\nu}(k)=[(i\epsilon_{n}+\mu-\bm{H}(\bm{k}))^{-1}]_{\mu\nu}$,
where $\mu$ is the chemical potential and $\bm{H}(\bm{k})$ is the matrix elements of the hopping term of the Hamiltonian in the momentum representation.
Here we abbreviate the wave numbers and the Matsubara frequencies as $k$ (for the fermionic case) or $q$ (bosonic).

Since we consider only $U_{\mu}$ as the interaction, introducing the ansatz,
\begin{equation}
U^{\rm{sp}}_{\mu\mu\mu\mu}=\frac{\ave{n_{\mu\uparrow}n_{\mu\downarrow}} }{\ave{n_{\mu\uparrow}} \ave{n_{\mu\downarrow}}}U_{\mu},
\end{equation}
susceptibilities can be determined from the following sum rules derived from the Pauli principle:
\begin{subequations}
\label{sum rules}
\begin{eqnarray}
-\frac{2T}{N}\sum_{q}\chi^{\rm{sp}}_{\mu\mu\mu\mu}(q) &=& n_{\mu}-2\ave{ n_{\mu\uparrow}n_{\mu\downarrow} } , \label{spin sum} \\
-\frac{2T}{N}\sum_{q}\chi^{\rm{ch}}_{\mu\mu\mu\mu}(q) &=& n_{\mu}+2\ave{ n_{\mu\uparrow}n_{\mu\downarrow} } - n_{\mu}^{2} , \label{charge sum}
\end{eqnarray}
\end{subequations}
where $n_{\mu}$ is the particle number per site of orbital $\mu$, obtained from $-\frac{T}{N}\sum_{k}G^{0}_{\mu\mu}(k)=n_{\mu}$.
However the ansatz introduced here violates the electron-hole symmetry. Therefore if $n_{\mu}>1$, considering the electron-hole transformation, 
the ansatz should be modified as
\begin{equation}
U^{\rm{sp}}_{\mu\mu\mu\mu}=\frac{\ave{n^{\rm{h}}_{\mu\uparrow}n^{\rm{h}}_{\mu\downarrow}} }
{\ave{n^{\rm{h}}_{\mu\uparrow}} \ave{n^{\rm{h}}_{\mu\downarrow}}}U_{\mu},
\end{equation}
where $n^{\rm{h}}_{\mu\sigma}=1-n_{\mu\sigma}$.
Since $n_{\mu}>1$ is satisfied for any band filling used in this study, we make use of the transformed ansatz.

Using the obtained susceptibilities as described above, the dressed Green's function $\bm{G}(k)$ is determined by Dyson equation:
\begin{equation}
\bm{G}(k)^{-1} = \bm{G}^{(0)}(k)^{-1} - \bm{\Sigma}(k) , \label{Dyson}
\end{equation}
and the self-energy $\bm{\Sigma}(k)$ is given by
\begin{eqnarray}
\Sigma_{ll^{\prime}}(k) &=& \frac{1}{2} \frac{k_{B}T}{N} \sum_{q} \left[ \bm{U}^{\rm{sp}}\bm{\chi}^{\rm{sp}}(q)\bm{U} \right. \nonumber \\
 &&\left. +\bm{U}^{\rm{ch}}\bm{\chi}^{\rm{ch}}(q)\bm{U}\right]_{lml^{\prime}m^{\prime}}G_{mm^{\prime}}^{(0)}(k-q) . \label{Self}
\end{eqnarray}
Solving linearized Eliashberg equation,
\begin{eqnarray}
\lambda \Delta_{ll^\prime}(k) &=& \sum_{k^\prime m_i} \Gamma^{s}_{l m_1 m_4 l^\prime}(k,k^\prime)  G_{m_1 m_2}(k^\prime)
\nonumber \\ && \times \Delta_{m_2 m_3}(k^\prime) G_{m_4 m_3}(-k^\prime) , \label{Eliash}
\end{eqnarray}
the eigenvalue $\lambda$ and the anomalous self-energy $\bm{\Delta}(k)$ are obtained. Here the singlet pairing interaction $\bm{\Gamma}^{s}(q)$ is given by
\begin{equation}
\bm{\Gamma}^{s}(q) = -\bm{U} - \frac{3}{2} \bm{U}^{\rm{sp}} \bm{\chi}^{\rm{sp}}(q) \bm{U} +  \frac{1}{2} \bm{U}^{\rm{ch}} \bm{\chi}^{\rm{ch}}(q) \bm{U},
\end{equation}
where $\bm{U}$ is the interaction matrix for the bare vertex. The 
superconducting transition temperature $T_c$ is the temperature where $\lambda$ 
reaches unity. In the present study, we calculate $\lambda$ at a fixed temperature and use it as a measure for $T_c$.

\begin{figure}[t]
\centering
\includegraphics[width=7.5cm,clip,bb=0 0 600 500]{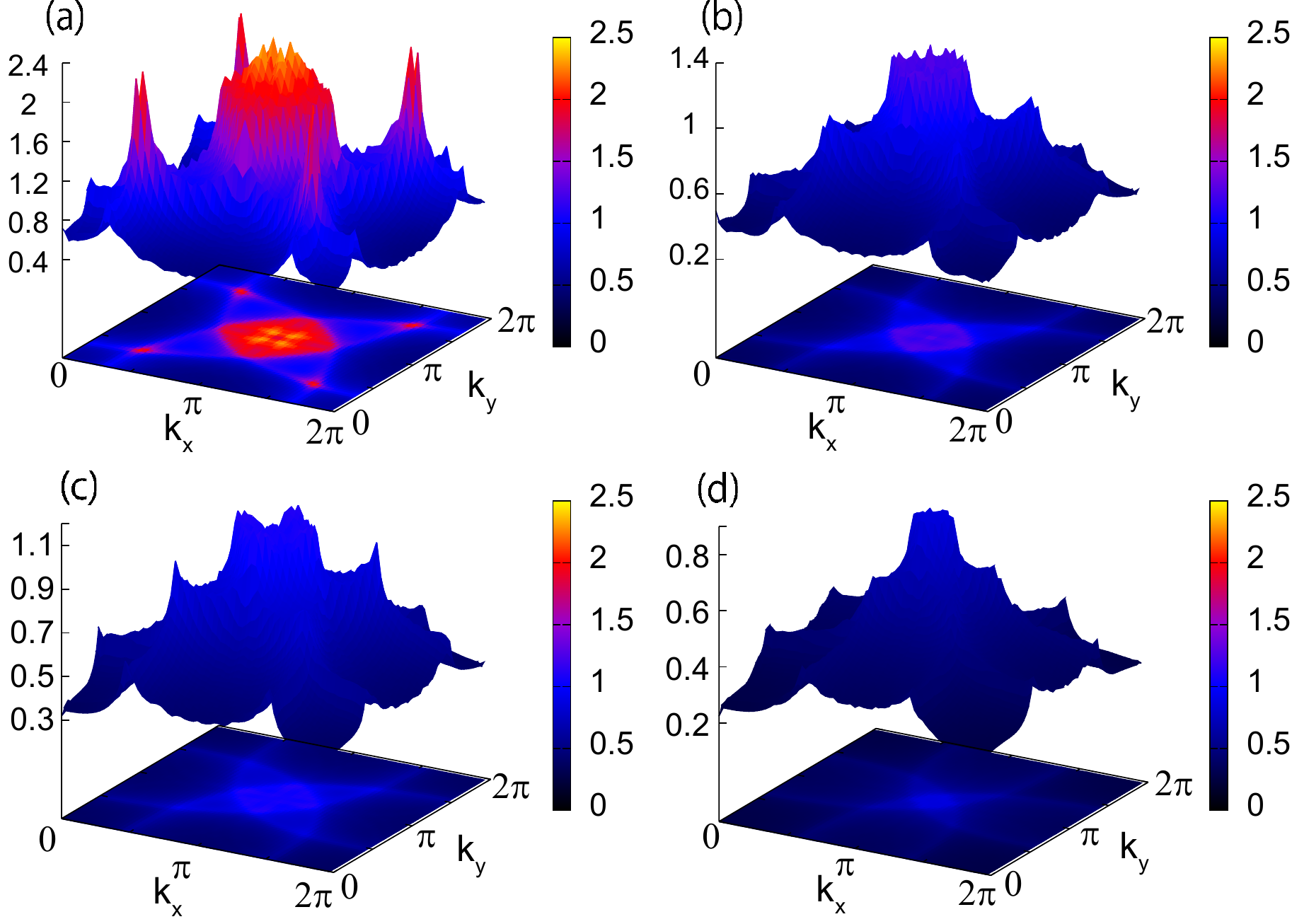}
\caption{(color online). Spin susceptibility of the three-band models $\sum_{\mu}\chi^{\rm{sp}}_{\mu\mu\mu\mu}(\bm{k},\omega=0)$.(a)Hg system, band filling $n=4.85$,
		(b)Hg system, band filling $n=5.0$, (c)Nd system,band filling $n=5.0$, (d)Nd system, band filling $n=5.15$.}
\label{fig3}
\end{figure}
\begin{figure}[t]
\centering
\includegraphics[width=8.0cm,clip,bb=0 0 700 500]{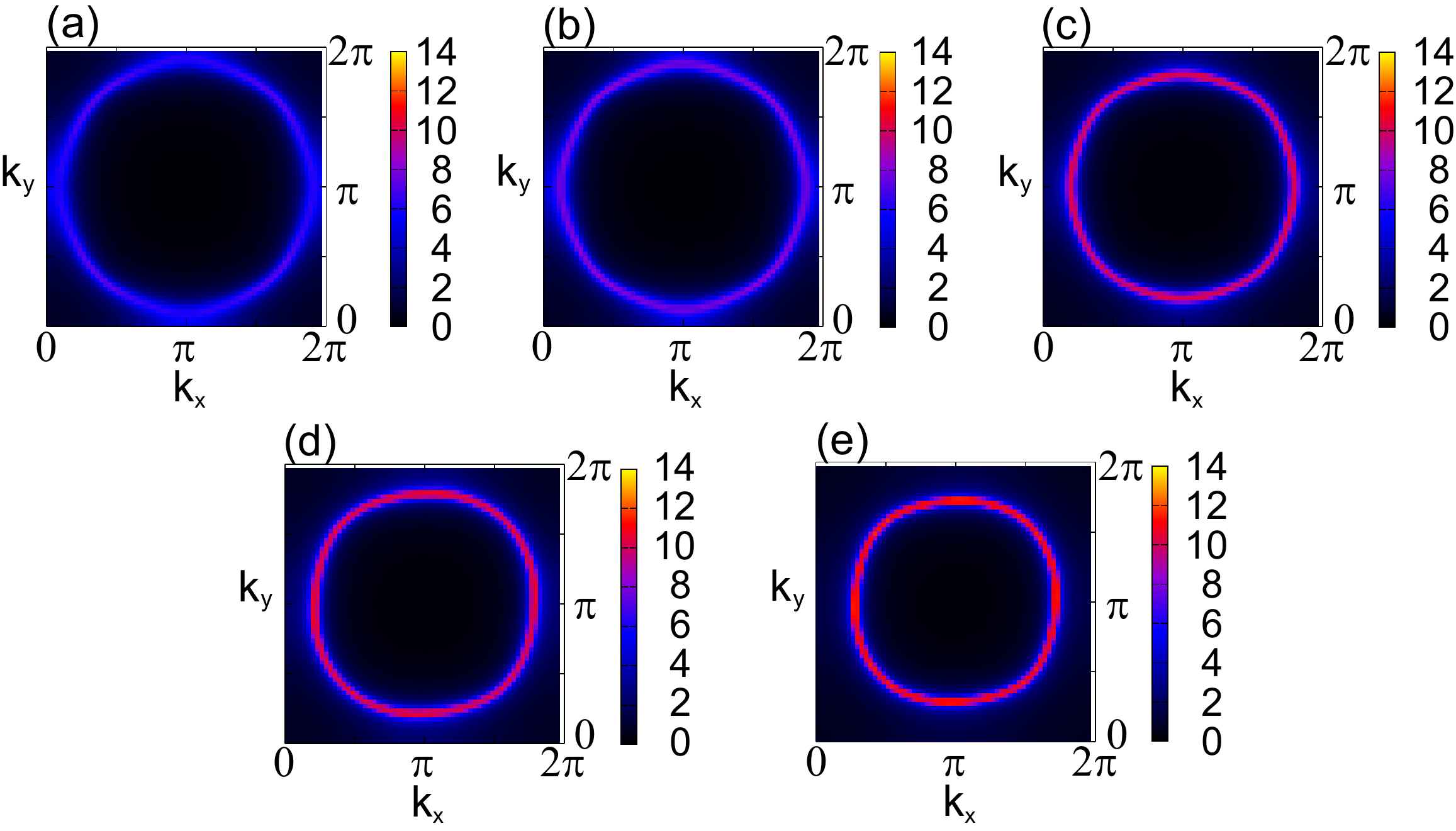}
\caption{(color online). Absolute value of the dressed Green's function of the three-band models $\abs{G_{dd}(\bm{k},i\epsilon_{n=0})}$. 
		(a)Hg system, band filling $n=4.80$, (b)Hg system, band filling $n=4.85$, (c)Hg system, band filling $n=5.0$, 
		(d)Nd system, band filling $n=5.0$, (e)Nd system, band filling $n=5.15$.}
\label{fig4}
\end{figure}

Let us move on to the calculation results of the effective three-band model for HgBa$_2$CuO$_4$ and Nd$_2$CuO$_4$.
The spin susceptibility $\sum_{\mu}\chi_{\mu\mu\mu\mu}^{\rm{sp}}(\bm{k},\omega=0)$ 
and the absolute value of the dressed Green's function $\abs{G_{dd}(\bm{k},i\epsilon _{n=0})}$ are shown in Fig.\ref{fig3} and Fig.\ref{fig4}, respectively.
As the number of electrons decrease from the electron-doped region $(n>5)$ to the hole-doped region $(n<5)$, 
peaks of the spin susceptibility around $(\pi,\pi)$ and the $\Gamma$ point are enhanced and the absolute value of the dressed Green's function is suppressed.
It can be seen in Fig.\ref{fig3}(a)(b) that the Green's function is particularly suppressed around $(\pi,0)$ $(0,\pi)$, namely the hot spots (see the figures in the supplementary material\cite{supplementary_ogura2015}, in which the hot spots are more clearly visible).

These behaviors can be explained as a combined effect of 
the following two factors.
First, since the $d_{x^2-y^2}$ orbital is not half-filled due to the $d$-$p$ hybridization,  the $d_{x^2-y^2}$ orbital approaches the half-filling by decreasing the number of electrons.
Because of this, to satisfy the sum rule for the spin susceptibility  $\chi^{\rm{sp}}_{dddd}(q)$ within the $d$ orbital, the vertex $U_{dddd}^{\rm{sp}}$ necessarily increases. Therefore the spin susceptibility increases with decreasing the number of electrons. Secondly, since the Fermi level approaches the van Hove singularity point of the band structure as the number of electrons is reduced, the spin susceptibility around the $\Gamma$ point is enhanced. The enhancement of the spin fluctuation results in the increase of the self energy, which in turn suppresses the Green's function.

\begin{figure}[t]
\centering
\includegraphics[width=8.0cm,clip,bb=0 0 650 500]{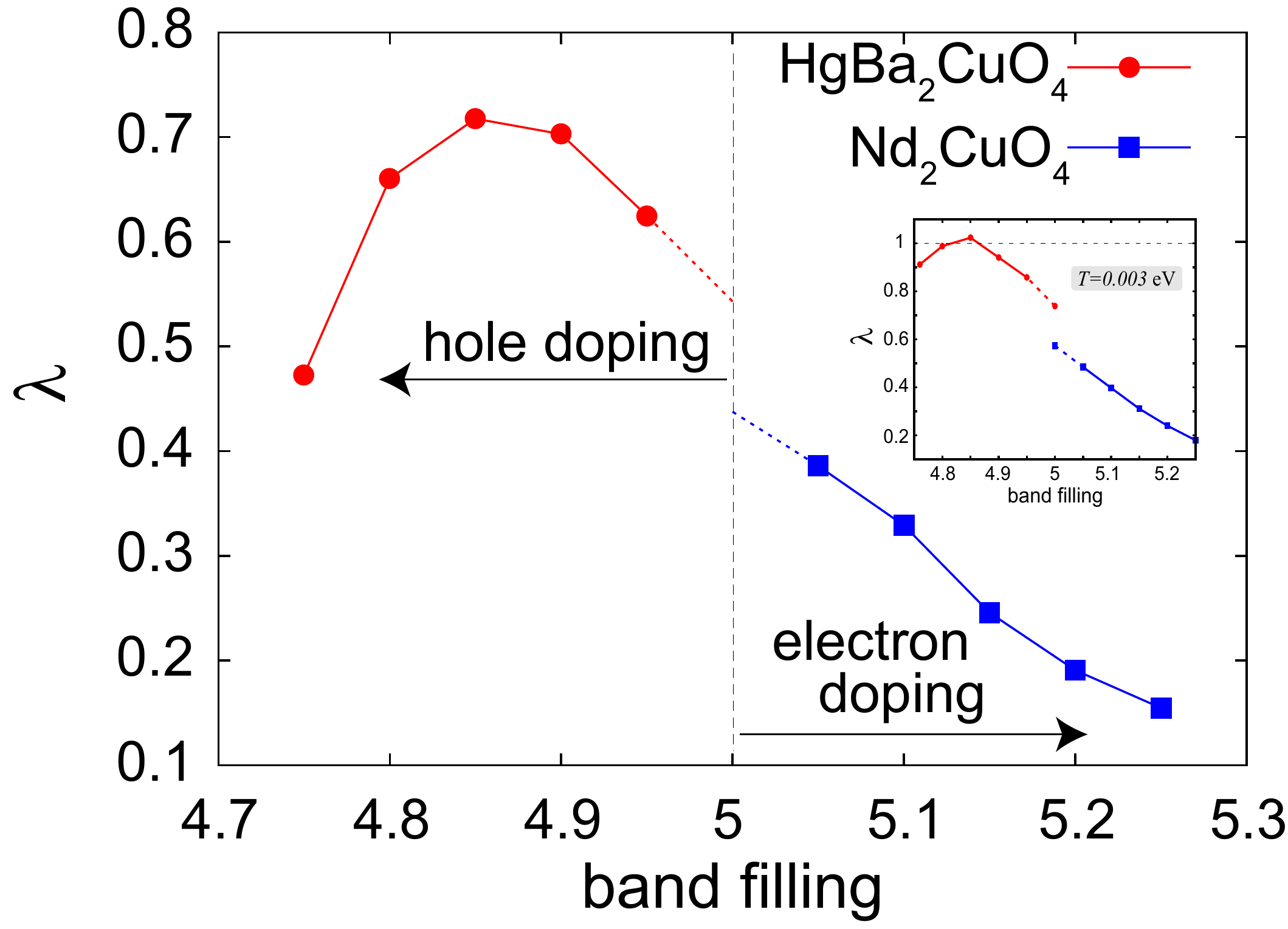}
\caption{(color online). The doping dependence of the $d$-wave eigenvalue of the linearized Eliashberg equation $\lambda$ in the three-band model.
	The inset shows the doping dependence of $\lambda$ at a lower temperature $T=0.003$eV.}
\label{fig5}
\end{figure}

We show the band filling dependence of the $d$-wave pairing eigenvalue $\lambda$ of the linearized Eliashberg equation (measure of $T_c$) in Fig.\ref{fig5}.
This result is consistent with the doping dependence of $T_c$ in both the electron- and the hole-doped region (except near the non-doped regime, which we will discuss later).
As shown in the inset of Fig.\ref{fig5}, this feature remains at the temperature where the $d$-wave eigenvalue $\lambda$ is above unity near the optimal doping rate ($T=0.003\mathrm{eV}$, 80$\times$80 $k$-point meshes, and 8192 Matsubara frequencies).
This result can be interpreted as follows.
Monotonic increase of $T_c$ in the electron-doped region as the number of electrons is reduced arises from the enhancement of $\chi^{\rm{sp}}_{dddd}(q)$ around $(\pi,\pi)$, which works in favor of the $d$-wave pair scattering.
As the band filling enters the hole-doped region, $\chi^{\rm{sp}}_{dddd}(q)$ is further enhanced around $(\pi,\pi)$, so that $\lambda$ also increases.
However, both the suppression of the Green's function and the enhancement of $\chi^{\rm{sp}}_{dddd}(q)$ around the $\Gamma$ point work against $d$-wave superconductivity, and therefore $\lambda$ turns to decrease with further hole doping beyond $\delta_h=0.15$, where $\delta_h=5-n$ is the hole doping rate.
Thus, the doping dependence of the superconducting transition temperature is naturally understood  in both the electron- and the hole-doped cases.

Since the Mott transition is not described within the formalism used in the present study, in Fig.\ref{fig5} we show the result near $n=5$ by dashed lines (the calculations have been done also at $n=5$ nonetheless).
The absence of the insulating state in the 
non-doped case is attributed to the insufficiency of  the evaluation of the local electron correlation effects. The inclusion of further 
electron correlation effects is left for future study. Nonetheless, we can 
expect that the inclusion of such effects will probably make the dome-like 
feature in the hole-doped region more prominent, while it should somewhat 
reduce the enhancement of $\lambda$ in the underdoped regime of the electron-doped case, which seems rather strong in the present calculation compared to experimental observations\cite{Krockenberger2013,Brinkmann1995}. 
Hence, the inclusion of further local correction is likely to make the doping dependence of $T_c$ even closer to those observed experimentally. 
Another related issue is the pseudogap problem in the underdoped regime. This has been addressed by TPSC in ref.\cite{Vilk1997} for the single band model, but the situation can be different in the case of the three-band model with realistic band structure. This also serves as an interesting future problem.

To summarize, we have studied the doping dependence of superconductivity 
for the three band model of Nd$_2$CuO$_4$ and HgBa$_2$CuO$_4$ using the TPSC 
method. The eigenvalue of the Eliashberg equation $\lambda$ exhibits an optimal doping around $n=4.85$ (hole concentration $\delta_h=0.15$) 
in the hole-doped region and 
varies monotonically in the electron-doped region, consistent with the experiment.
It is found to be understood naturally in terms of the electron-hole asymmetry due to the $d$-$p$ hybridization and the band-filling-dependent vertex correction.

Part of the numerical calculations were performed
at the facilities of the Supercomputer Center,
Institute for Solid State Physics, 
University of Tokyo.
This study has been supported by 
Grants-in-Aid  for Scientific Research No.26247057 
from the Japan Society for the Promotion of Science.
%
%

%

%

%
\bibliography{ronbun}

\end{document}